# K-shell decomposition reveals hierarchical cortical organization of the human brain

**Nir Lahav*[1], Baruch Ksherim*[1], Eti Ben-Simon[2,3], Adi Maron-Katz[2,3], Reuven Cohen[4], Shlomo Havlin[1].**

*1. Dept. of Physics, Bar-Ilan University, Ramat Gan , Israel*
*2.Sackler Faculty of Medicine, Tel Aviv University, Tel-Aviv , Israel*
*3.Functional Brain Center, Wohl Institute for Advanced Imaging, Tel-Aviv Sourasky Medical Center, Tel- Aviv, Israel*
*4.Dept. of Mathematics, Bar-Ilan University, Ramat Gan, Israel*

**Authors contributed equally to this work*

**Corresponding author:** Nir Lahav, Dept. of Physics, Bar-Ilan University, Ramat Gan, Israel. E-mail: freenl@gmail.com

**Abstract**

In recent years numerous attempts to understand the human brain were undertaken from a network point of view. A network framework takes into account the relationships between the different parts of the system and enables to examine how global and complex functions might emerge from network topology. Previous work revealed that the human brain features 'small world' characteristics and that cortical hubs tend to interconnect among themselves. However, in order to fully understand the topological structure of hubs, and how their profile reflect the brain's global functional organization, one needs to go beyond the properties of a specific hub and examine the various structural layers that make up the network.

To address this topic further, we applied an analysis known in statistical physics and network theory as **k-shell decomposition analysis.** The analysis was applied on a human cortical network, derived from MRI\DSI data of six participants. Such analysis enables us to portray a detailed account of cortical connectivity focusing on different neighborhoods of inter-connected layers across the cortex. Our findings reveal that the human cortex is highly connected and efficient, and unlike the internet network contains no isolated nodes. The cortical network is comprised of a nucleus alongside shells of increasing connectivity that formed one connected giant component, revealing the human brain's global functional organization. All these components were further categorized into three hierarchies in accordance with their connectivity profile, with each hierarchy reflecting different functional roles. Such a model may explain an efficient flow of information from the lowest hierarchy to the highest one, with each step enabling increased data integration. At the top, the highest hierarchy (the nucleus) serves as a global interconnected collective and demonstrates high correlation with consciousness related regions, suggesting that the nucleus might serve as a platform for consciousness to emerge.

**"..And you ask yourself, where is my mind?" The pixies (Where is my mind)**

## *Introduction*

The human brain is one of the most complex systems in nature. In recent years numerous attempts to understand such complex systems were undertaken, in physics, from a network point of view (Carmi, 2007; Cohen and Havlin, 2010; Newman, 2003; Colizza and Vespignani, 2007; Goh *et al.*, 2007). A network framework takes into account the relationships between the different parts of the system and enables to examine how global and complex functions might emerge from network topology. Previous work revealed that the human brain features 'small world' characteristics (i.e. small average distance and large clustering coefficient associated with a large number of local structures (Achard *et al.*, 2006; Bullmore and Sporns, 2009; He *et al.*, 2007; Ponten *et al.*, 2007; Reijneveld *et al.*, 2007; Sporns *et al.*, 2004; Sporns and Zwi, 2004; Stam *et al.*, 2007; Stam and Reijneveld, 2007; van den Heuvel *et al.*, 2008)), and that cortical hubs tend to interconnect and interact among themselves (Achard *et al.*, 2006; Buckner *et al.*, 2009; Eguiluz *et al.*, 2005; van den Heuvel *et al.*, 2008). For instance, van den Heuvel and Sporns demonstrated that hubs tend to be more densely connected among themselves than with nodes of lower degrees, creating a closed exclusive "*rich club*" (Collin *et al.*, 2014; Harriger *et al.*, 2012; van den Heuvel and Sporns, 2011; van den Heuvel *et al.*, 2013). These studies, however, mainly focused on the individual degree (i.e. the number of edges that connect to a specific node) of a given node, not taking into account how their neighbors' connectivity profile might also influence their role or importance. In order to better understand the topological structure of hubs, their relationship with other nodes, and how their connectivity profile might reflect the brain's global functional organization, one needs to go beyond the properties of a specific hub and examine the various structural layers that make up the network.

In order to explore the relations between network topology and its functional organization we applied a statistical physics analysis called **k-shell decomposition** (Adler, 1991; Alvarez-Hamelin *et al.*, 2005a; Alvarez-Hamelin *et al.*, 2005b; Carmi, 2007; Modha and Singh, 2010; Pittel *et al.*, 1996; Garas *et al.*, 2010) on a human cortical network derived from MRI and DSI data. Unlike regular degree analysis, k-shell decomposition does not only check a node's degree but also considers the degree of the nodes connected to it. The k-shell of a node reveals how central this node is in the network with respect to its neighbors, meaning that a higher k-value signifies a more central node belonging to a more connected neighborhood in the network. By removing different degrees iteratively, the process enables to uncover the most connected area of the network (i.e., the **nucleus**) as well as the connectivity **shells** that surround it. Therefore, every shell defines a neighborhood of nodes with similar connectivity (see Fig. 1). A few studies have already applied this analysis in a preliminary way, focusing mainly on the network's nucleus and its relevance to known functional networks (Hagmann *et al.*, 2008; van den Heuvel and Sporns, 2011). For instance, Hagmann et al. revealed that the nucleus of the human cortical network is mostly comprised of *default mode network* regions (Hagmann *et al.*, 2008). However, when examined more carefully, k-shell decomposition analysis, as shown here, enables the creation of a topology model for the entire human cortex taking into account the nucleus as well as the different connectivity shells ultimately uncovering a reasonable picture of the global functional organization of the cortical network. Furthermore, using previously published k-shell analysis of internet network topology (Carmi, 2007) we were able to compare cortical network topology with other types of networks.

We hypothesize that using k-shell decomposition would reveal that the human cortical network exhibits a *hierarchical structure* reflected by shells of higher connectivity, representing increasing levels of data processing and integration all the way up to the nucleus. We further assume that different groups of shells would reflect various cortical functions, with high order functions associated with higher shells. In this way we aim to connect the structural level with the functional level and to uncover how complex behaviors might emerge from the network.

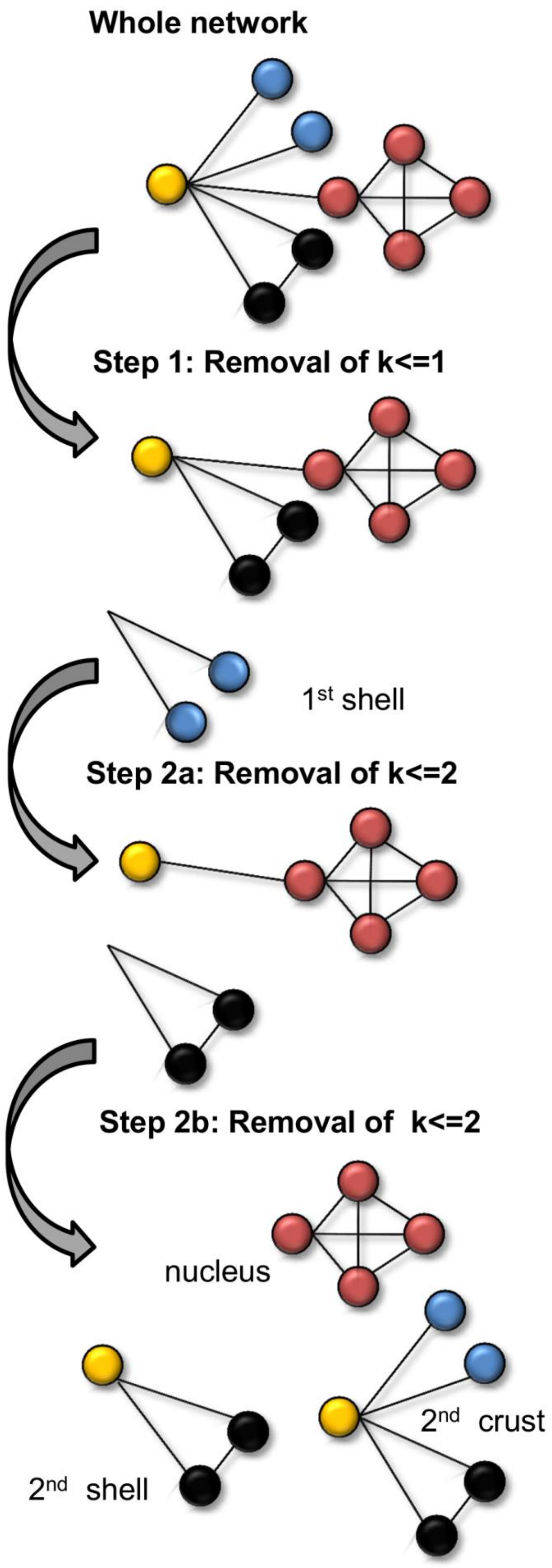

**Figure 1: K-shell decomposition process.**

K-shell decomposition takes into account the degree of the node as well as the degree of the nodes connected to it.

This example shows the difference of the K-shell method compared with regular degree count. Top panel: The whole network. The yellow node is a hub (k=5) and thus one might think that it would be in the nucleus. But on the first step of the process (k=1), two of its neighbors will be removed to the first shell (blue). When re-computing the degree of the remaining nodes we notice there are no more nodes with only one link. The remaining network is the 1st core. On step 2a (k=2), another two of its neighbors will be removed (black). Then, when re-computing the degree of each node (step 2b), the yellow node has a low degree (k=1) and will be removed to the second shell. The process stops in k=3 when the remaining nodes will be removed and no node will remain in the network. K-core is composed of the remaining network in a given k step and the **nucleus** is defined as the final k-core in the process. The nucleus of this network is thus the 2nd-core, the group of the last remaining nodes (red). K-crust includes the nodes that have been removed until step k of the process. This network has 5 nodes in its 2nd-crust (blue, black and yellow. for more details see methods).

## Materials and methods

### *Imaging*

The networks for our analysis were derived from two combined brain imaging methods, MRI/DSI recorded by **Patric Hagmann's group** from University of Lausanne (for all the functions and data sets, please refer to : http://www.brain-connectivity-toolbox.net/). Using this data, clusters of gray matter formed the nodes while fibers of white matter formed the edges of the cortical network. In this technique, 998 cortical ROIs were used to construct the nodes of each network and 14,865 edges were derived from white matter fibers (for more specific details please see Hagmann P. et al. (Hagmann *et al.*, 2008)). Six structural human cortical networks were transformed into six connection matrices by Patric Hagmann's group, derived from five right handed subjects (first two networks were derived from the same subject in different times). These connection matrices were utilized to calculate the network's properties and to apply the k-shell decomposition analysis. We used binary connection matrices ('1' –connected, '0' –disconnected) and not weighted connection matrices because of known difficulties in determining the appropriate weights and how to normalize them (Hagmann *et al.*, 2007; Hagmann *et al.*, 2003; Van Den Heuvel and Pol, 2010; van den Heuvel and Sporns, 2011). In order to connect between our structural network and known functional networks the 998 nodes were clustered into 66 known anatomical regions in accordance with Hagmann et al. (Hagmann *et al.*, 2008).

### *Network theory*

Several network characteristics were used in our analysis (see supplementary material 6 for further details):

*Degree (k)* of a node is the number of edges that connect to the node.

*Hub* is a node with degree above the average degree of the network.

*Distance* between nodes is the shortest path between node i and node j.

*Average diameter (L)* of the network is denoted by:

$$L = \frac{1}{N(N-1)} \sum_{i \neq j} d_{ij}$$

$d_{ij}$ **–** distance between node i and node j ***;*** $N$ **–** total number of nodes in the network

*Local clustering coefficient* ($c_i$) of a node i reflects the probability that "my friend's friend will also be my friend" (computed for each node). **Clustering coefficient** is the average over all local $\mathbf{c_i}$ and it provides estimation of the amount of local structures in the network. Topologically it means that the network will have a large quantity of triangles: $C = \frac{1}{N}\sum_i c_i$

*Small-world networks* are networks that are significantly more clustered than random networks, yet have approximately the same characteristic path length as random networks (high clustering coefficient and low average distance).

*Assortativity coefficient* is the Pearson correlation coefficient of degree between pairs of linked nodes. Positive values indicate a correlation between nodes of similar degree, while negative values indicate relationships between nodes of different degree**.** Assortativity coefficient lies between −1 and 1.

We also examined whether the cortical network exhibits a **hierarchal structure** (not to be confused with the hierarchies derived from k-shell decomposition analysis) in which hubs connect nodes which are otherwise not directly connected. Networks with a hierarchal structure have a power law clustering coefficient distribution- $C \sim K^{-\beta}$ which means that as the node degree increases (k) the clustering coefficient (C) decreases. The presence of hubs with low clustering coefficient means that the network has a hierarchal structure (since hubs connect nodes which are not directly connected, triangles with hubs are not frequent).

*Module structures*: the network's modular structure (community structure), is revealed by subdividing the network into groups of nodes, with a maximally possible number of within group links, and a minimally possible number of between-group links.

*K-shell decomposition method*

In the k-shell decomposition method we revealed the network's nucleus as well as the shells that surround it. The k-shell of a node indicates the centrality of this node in the network with respect to its neighbors. The method is an iterative process, starting from degree k=1 and in every step raising the degree to remove nodes with lower or similar degree, until the network's nucleus is revealed, along the following steps:

**Step 1**. Start with connectivity matrix M and degree k=1.

**Step 2**. Remove all nodes with degree ≤ k, resulting in a pruned connectivity matrix M′.

**Step 3**. From the remaining set of nodes, compute the degree of each node. If nodes have degree ≤ k, step 2 is repeated to obtain a new M′; otherwise, go back to step 1 with degree k=k+1 and M=M′.

**Stop** when there are no more nodes in M′ (M′=0).

The k-shell is composed of all the new removed nodes (along with their edges) in a given k step. Accumulating the removed nodes of all previous steps (i.e. all previous k-shells) is termed the k-crust. The k-core is composed of the remaining network in a given k step and the **nucleus** is defined as the final k-core in the process. In the end of every step a new k-shell, k-crust and k-core are produced of the corresponding k degree. In the end of the process the nucleus is revealed with the most central nodes of the network, and the rest of the nodes are removed to the different **shells** (see Fig. 1). Typically, in the process of revealing the nucleus, all removed nodes in the k-crust eventually connect to each other forming one *giant component*.

The uniqueness of k-shell decomposition method is that it takes into account both the degree of the node as well as the degree of the nodes connected to that node. Thus, we can examine groups of nodes, every group with its own unique connectivity pattern. In this way one can examine cortical anatomical regions according to their connectivity neighborhood. For each node in the network we determined its *shell level* (i.e. to which shell it belongs, or if it survived the whole process, it belongs to the highest level – the nucleus). We then calculated *shell levels* for every anatomical region, comprised of many nodes, according to the weighted average *shell level* of its nodes.

***Statistics and random networks***

In order to evaluate the significant of the properties of the cortical network each result was compared to that of a randomized network. The network was randomized by keeping the degree distribution and sequence of the matrix intact and only randomizing the edges between the nodes (Rubinov and Sporns, 2010). For each cortical network several random networks were computed with different amount of randomized edges (from 1% until 100% of the edges). This process was repeated several times iteratively. K-shell decomposition was applied for each of the randomized networks. Since the results of the cortical network were resilient to small perturbations (1% of the edges randomized) we raise the amount of randomization. For greater amount of randomization the results were fixed around an average value after 5 iterations (or more) using 100% random edges. Thus we took the random networks to be with 100% randomized edges and 5 iterations.

To assess statistical significance of our results across networks, permutation testing was used (Van Den Heuvel and Pol, 2010). Matrix Correlations across 6 networks were computed and compared with correlations obtained from 1000 random networks. These random network correlations yielded a null distribution comprised of correlations between any two networks obtained from the random topologies. Next, we tested whether the real correlations significantly exceeded the random correlations, validated by a p-value< 0.01. Moreover, the significance of the observed connectivity within and between hierarchies was evaluated using a random permutation test. In this test, each node was randomly assigned with a hierarchy, while preserving the connectivity structure of the graph as well as hierarchy sizes. This process was repeated 10,000 times, and in each repetition, the number of connections within each hierarchy and between each pair of hierarchies was recorded. For each pair of hierarchies, a connectivity p-value was calculated using the fraction of the permutations in which the number of connections linking them was equal or higher than this number in the real data. Resulting p-values were corrected for multiple comparisons using the false discovery rate (FDR) procedure thresholded at 0.05.

## *Results*

### *Cortex network topology*

The results of the K-shell decomposition process revealed that the human cortex topology model has an "egg-like" shape (see Fig. 2). In the "middle", 22% (± 12%) of the networks' nodes formed the nucleus ("the yolk" in the egg analogy) and "surrounding" the nucleus about 77% (± 12%) of the removed nodes formed the shells. These removed nodes did not reach the nucleus and connected to each other to form one *giant component*. The nucleus has on average 217 nodes (± 117) and the giant component has on average 770 nodes (± 121). The rest of the nodes are *isolated nodes*. These removed nodes did not connect to the giant component, and essentially connect to the rest of the network solely through the nucleus (some nodes are not connect to any other node in the network and thus were removed; on average 9±6 nodes per cortical network).

Over all 6 networks, the average k-core of the nucleus was 19(±1), which means that during the iterative process the nucleus was revealed after the removal of 19(±1) shells. Thus, the minimum degree in the nucleus is 20 and the average degree of the nodes in the nucleus is 45 (±4). In comparison, the average degree across the entire cortical network is 29 (±1), demonstrating that the nucleus contains hubs with significantly higher degree than that of the average network. In addition, the nucleus had considerably lower average distance compared to the average distance of the entire cortical network (2 ±0.2 vs. 3±0.1 , respectively). This finding means that it takes 2 steps, on average, to get from one node to any other node in the 217 nodes of the nucleus.

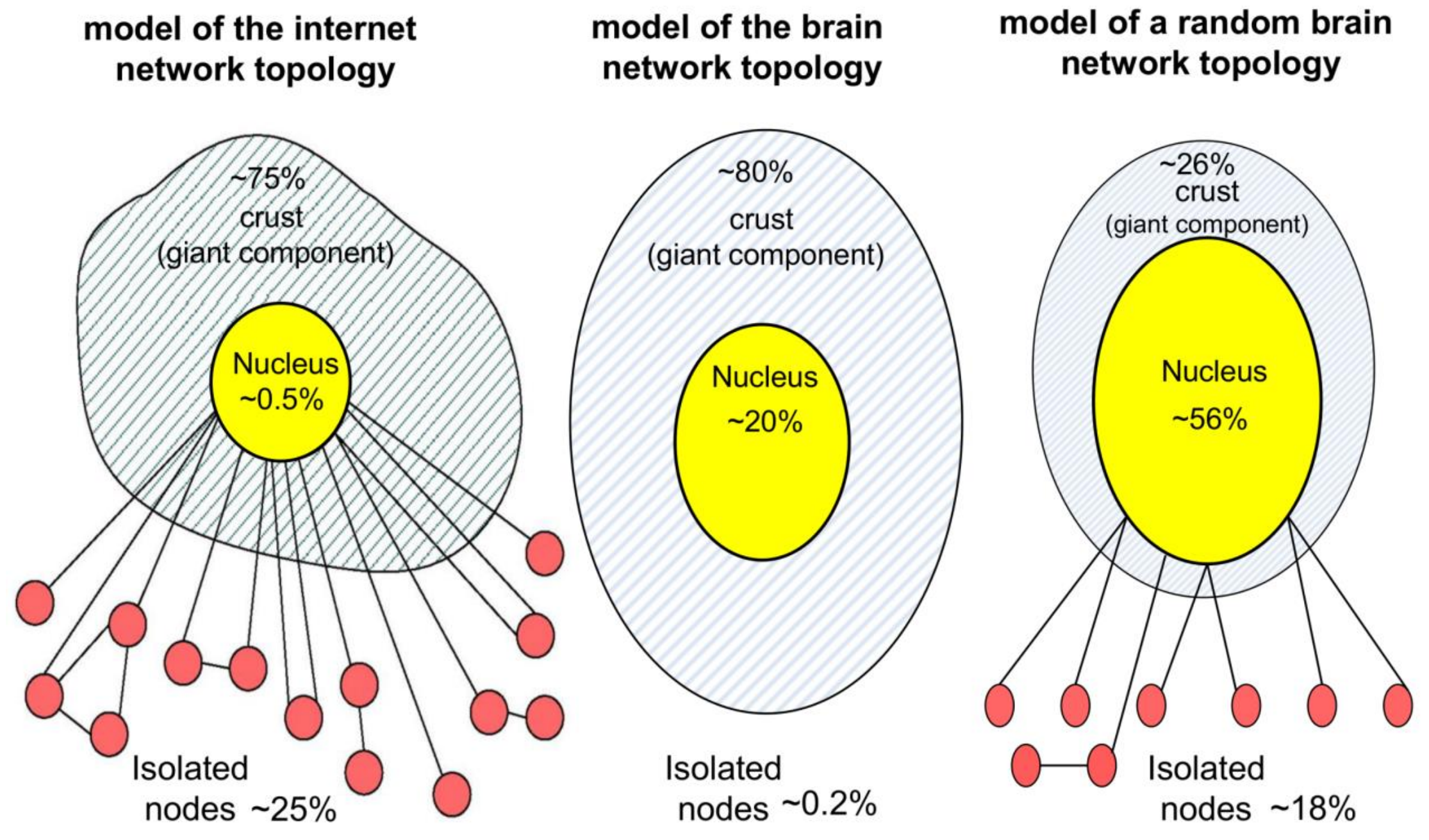

**Fig. 2: Topology of the cortical network.**
Topology of the cortical network (middle) compared with the internet topology, after Carmi et al(Carmi, 2007) (left) and random cortex network (right). In the cortical network the nucleus consists of 20% of the nodes while the remaining 80% compose a one giant component from all the removed nodes in the different shells. Note, a much bigger nucleus in the random cortical network and contrary to the cortical network larger amount of isolated nodes in both random and internet topologies.

The giant component is formed in a process similar to a first order phase transition with several critical points, as for the internet (Carmi, 2007; Pittel *et al.*, 1996). In the beginning of the process islands of removed nodes were forming and growing, but at some stage all of these islands connect together to form the giant component (see Fig. S1 for more details). This abrupt phase transition occurred, on average, in k-crust 15 ±1 (i.e. big islands of removed nodes were formed in crust 14, comprised of all previous shells including shell 14, but in crust 15 all of these islands disappear and a single giant component is formed). There is no significant difference between the number of removed nodes that were added to crust 15 compared to crust 14, yet a phase transition had occurred, suggesting that the difference is in the amount of the removed hubs. In crust 15, for the first time, enough hubs (which connect to lower degree nodes) were removed at once and connect all the islands to form the giant component. Later, another critical point is observed. On average in crust 18 (±1), a very large amount of nodes are removed at once to join the giant component (on average 282 nodes comprising 37% ±10% of all the nodes in the giant component (see Fig. S1). This may suggest that the process reached yet another group of higher hubs which have been removed along with their connections. These hubs connect to significantly more nodes than the previous hubs leading to a massive removal of nodes. We also note that the giant component features small world characteristics similar to the entire network (C=0.4 for both giant component and the whole network, average distance is 3.6 ±0.5 for the giant component, slightly higher than that of the whole network (3±0.1), see Fig. S2).

*Cortex network topology in comparison to other networks*

The cortical network topology is found to be very different from the topologies of a randomized cortex or the internet network (at the autonomous systems level) which displayed a "medusa-like" shape (Carmi, 2007) (see Fig. 2). In addition to the nucleus and the giant component both random and internet topologies have a large amount of isolated nodes, forming the "medusa legs" in the medusa shape (on average 17% in the randomized cortical networks and 25% in the internet network, unlike close to 0.3% ±0.3% in the cortical network).

In addition, the average nucleus size of the randomized cortex is nearly three times bigger than the average nucleus of the human cortex (56% vs. 20%). The cortical nucleus contains only 50% of the hubs, the rest fall on average in the last 4-5 shells before the nucleus, while in the random cortex 100% of all hubs reached the nucleus (see Fig. S3). A network that displays a significant amount of hubs on several levels and not just in the nucleus could support a hierarchical structure that enables modular integration, as evident in cortical function (Bassett *et al.*, 2008; Christoff and Gabrieli, 2000; Gray *et al.*, 2002; Northoff and Bermpohl, 2004; Northoff *et al.*, 2006). Note that in the cortical network the hubs outside the nucleus start on average at shell 14-15 which supports the hypothesis that the first phase transition (shell $15 \pm 1$ ) is due to the removal of those hubs (as mentioned above).

***Correlation between topology and known brain functions***

In the k-shell decomposition analysis the connections of a node as well as its neighborhood determine at which shell that node will be removed. Neighborhood of High degree will be removed in a higher shell, or might survive the entire process and be part of the nucleus. Therefore, the giant component is comprised of different shells which represent different neighborhood densities of connectivity. These shells, corresponding to known cortical networks, enable an effective examination of cortical hierarchical organization.

We, therefore, examined the functional attributes of the nodes found in the nucleus and in all shells, by checking the shell level of every anatomical region (mapping how many nodes from the anatomical region have been removed to the different shells). Subsequently, we were able to score each anatomical region in accordance with its place in the network's hierarchy represented by its shell level. This characterization is demonstrated to be more accurate than just analyzing the average degree of each anatomical region (see Fig. S4 and supplementary material 1 for further details).

Furthermore, we examined the nucleus and revealed known functional areas that are **always** found in the nucleus across all 6 networks (see Fig. 3). These areas comprise the entire **bilateral midline region** and overlap with five major functional networks: motor and motor planning, the default network, executive control network, high order visual areas and the salience network (see Table 1 for full details). In contrast, several known functional areas were **never** in the nucleus across all 6 networks. These areas include most of the **right temporal lobe** (e.g. the fusiform gyrus, A1, V5), right Broca and Wernicke homologues and right inferior parietal cortex. Interestingly, all the areas that never appear in the nucleus are from the right hemisphere. Furthermore, 70% of all the lowest shells are from the right hemisphere while 60% of the areas that are **always** in the nucleus belong to the left hemisphere (see supplementary material 2 for more details ).

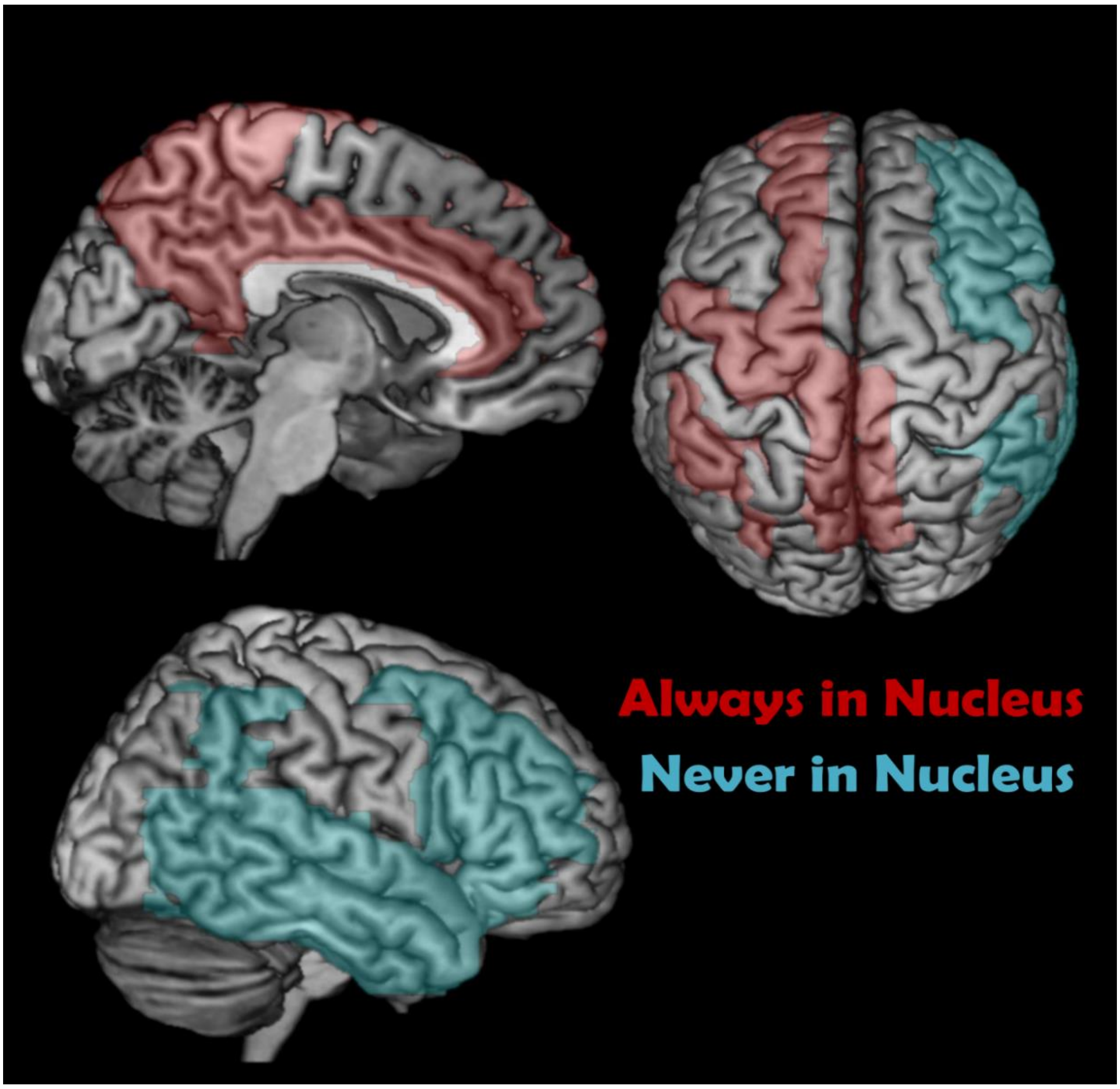

**Fig. 3: Anatomical regions and the network nucleus.** Brain maps displaying anatomical regions that are always in the nucleus (red) and never in the nucleus (blue). Note that all the regions that never reach the nucleus are from the right hemisphere.

Next, we used the critical points that were observed during the giant component formation (see supplementary material 3 for more details) in order to detect and establish different hierarchies of shells. Briefly, the creation of the giant component corresponded to the shell threshold of a middle hierarchy and the creation of the nucleus corresponded to the threshold of a high hierarchy. This analysis resulted in three major hierarchal groups (low, middle and high) as portrayed in Figure 4.

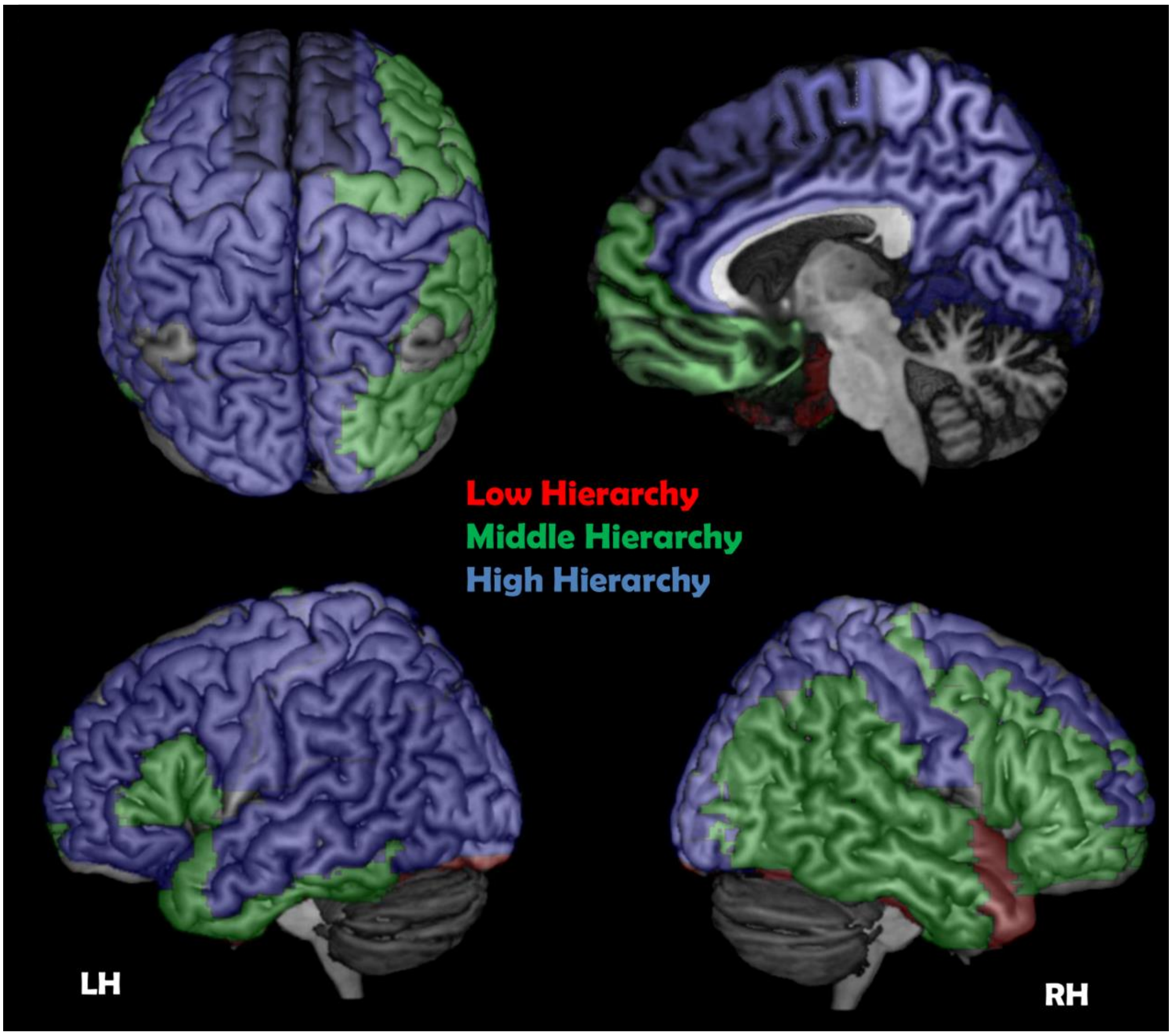

**Fig. 4: Anatomical regions according to their hierarchies.** Brain maps displaying cortical anatomical regions according to their hierarchies. Red – low hierarchy, green – middle hierarchy, blue – high hierarchy. One can divide the cortex to low hierarchy

regions found in the lateral bottom part of the cortex, middle hierarchy in the lateral middle part of the cortex, and high hierarchy in the lateral top and midline part of the cortex. RH - Right hemisphere, LH - Left hemisphere

The first hierarchal group consists of regions found in the lowest shells (average shell level 8.8, number of nodes/edges: 99/730 respectively). The removed nodes of this group are distributed across the shells with relatively high standard deviation (4.42, e.g. fusiform gyrus, entorhinal cortex, parahippocampal cortex. See Table 1 and Fig. 5 for full details). Notably, in this hierarchal group 75% of the regions were bilateral and 50% of the regions were never in the nucleus. The second hierarchy is a middle group which includes nodes found in the highest shells, but still not in the nucleus (number of nodes/edges: 335/4377 respectively). This group can be further subdivided to two subgroups, *distributed middle* and *localized middle* according to their average shell level and standard deviation. The average shell level of the distributed middle group is 14.5 (±3.07). This subgroup includes regions like right A1, right V5 and right Broca's homologue (for full details see table 1 and Fig. 5d). The average shell level of the localized middle group is 16.67 (±1.13). This subgroup includes regions like right wernicke homologue and right middle frontal gyrus. In the middle hierarchy 56% of the regions are bilateral and 40% of the regions are from the right hemisphere (in localized middle 88% right). 48% of the regions in this hierarchy were never found in the nucleus (for full details see Table 1 and Fig. 5c).

The third group is the highest hierarchy which contains regions predominantly found in the nucleus (number of nodes/edges: 561/8430 respectively). This group can also be subdivided to two subgroups, *distributed high* and *localized high* according to their average shell level and standard deviation. Average shell level of distributed high is 16.92 (±2.82). This subgroup includes the superior frontal gyrus, left Wernicke, left Broca and left V5. The average shell level of the localized high group is 19.30 (±0.97) and includes the precuneus and the cingulate cortex (for full details see Table 1 and Fig. 5). In this hierarchal group 69% of the regions were bilateral while 28% of the regions belonged to the left hemisphere. 44% of the regions in this hierarchy were always in the nucleus (66% in localized high). Altogether, all the regions that are always in the nucleus are from the high hierarchy while the regions that never reached the nucleus are from lower hierarchies.

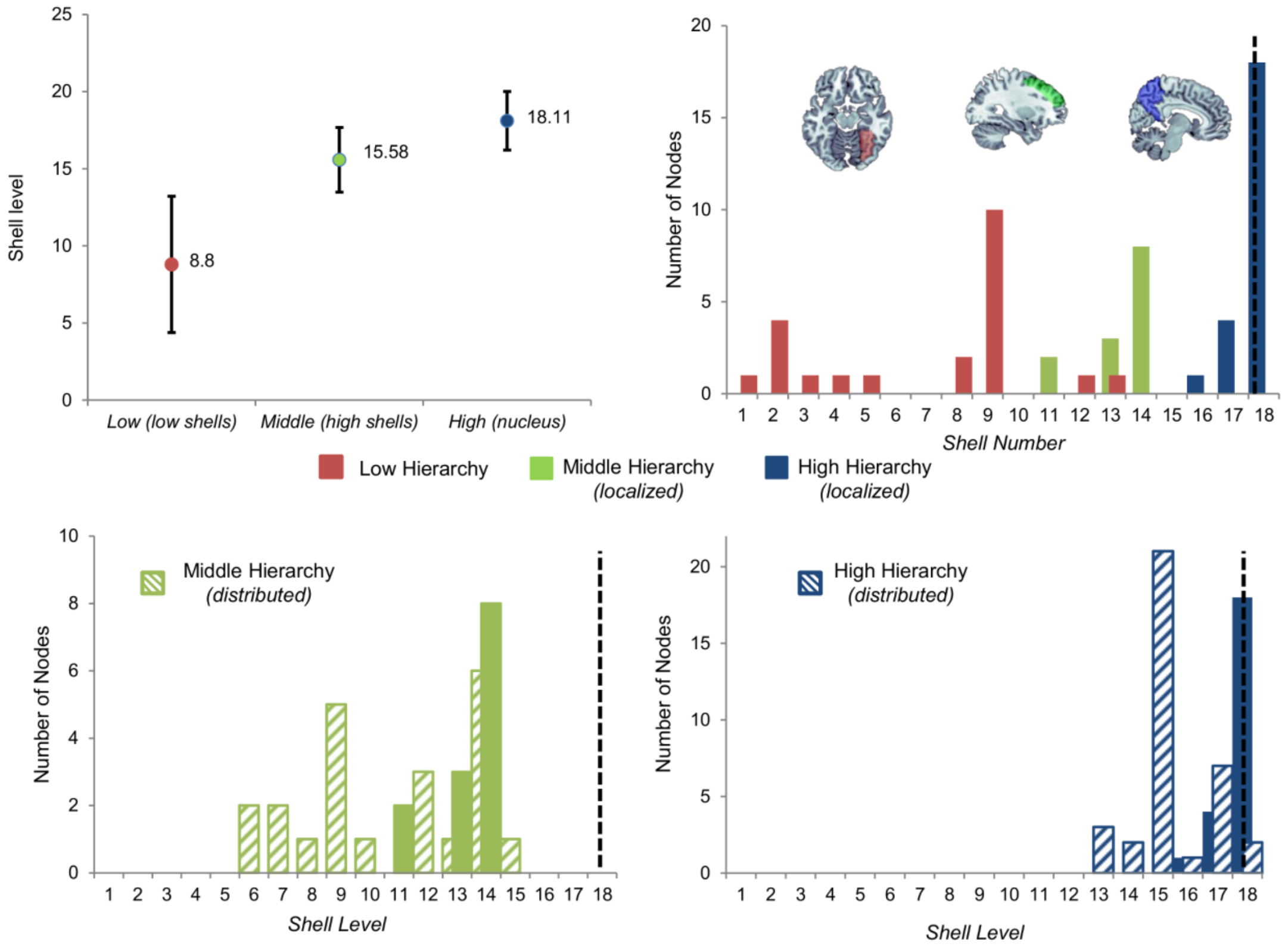

**Fig. 5: Hierarchies of the cortical network.** ***Top left panel:*** average shell level of the hierarchies. X-axis: hierarchy, Y-axis: shell level. ***Top right panel:*** an example of a single anatomical region representing each hierarchy (derived from average cortical network over all 6 networks. For exact data see supplementary data 1 and Fig. S5). Right Precuneus as an example of localized high hierarchy regions (blue). Notably, this area always reached the nucleus. Right caudal middle frontal as an example of localized middle hierarchy regions (green). Notably, this area never reached the nucleus. Right fusiform gyrus as an example of low hierarchy regions (red). Note the high standard deviation of the shell distribution. This region never reached the nucleus. X-axis: k-shell number; Y-axis: number of nodes. Dashed line: nucleus.

***Bottom left:*** Right lateral occipital cortex as an example of distributed middle hierarchy regions (striped green, localized middle hierarchy as above). X-axis: k-shell number; Y-axis: number of nodes. Dashed line: nucleus. ***Bottom right:*** Left precentral gyrus as an example of distributed high hierarchy regions (striped blue, localized high hierarchy as above). This area always reached the nucleus. X-axis: k-shell number; Y-axis: number of nodes, dashed line: nucleus.

Using the shell score we could further estimate the average shell level of known functional regions or networks (see Table S2). Interestingly, average shell level often reflected known functional lateralization as detailed in Table 2. For instance, while Broca's area is found in the nucleus, its right homologues never reached the nucleus. In a similar way, Wernicke's area is found in the high hierarchy and its right homologue in the middle hierarchy, again never reaching the nucleus. Right primary motor region and right TPJ are found in the middle hierarchy (and also never reached the nucleus) whereas their left counterparts are found in the high hierarchy (and left primary motor region always reached the nucleus). The functional network with the highest average shell level was the default mode network (DMN) with a score of 18.1. 81% of its regions were found in the high hierarchy with 70% always reaching the nucleus. Following the DMN, the salience and the sensorimotor networks also demonstrate high average shell level (17.3 and 17.5, respectfully) reflecting their high functional relevance. These results are detailed in Table 2 and in supplementary material 2.

*Connections between hierarchies*

In order to examine the connections between the different hierarchies, we compared the number of connections within each hierarchy to the number of connections with other hierarchies (calculated as a percentage of its total connections). Within the lowest hierarchy it was found that only 22% ±6.33% were self-connections and the rest were distributed between the middle group (30% ±3.36%) and the highest group (48% ±4.24%). In the middle hierarchy approximately half of the connections (52% ±2.6%) were self-connections and 41.5% ±2.6% were linked to the highest group. Interestingly, only 7% ±0.77% of the connections from the middle hierarchy were linked to the lowest hierarchy. The highest hierarchy exhibited the highest levels of self-connections (72% ±1.6%). Only 22.5% ±1.5% of its connections were linked to the middle hierarchy and 6% ±0.6% to the lowest hierarchy (for more details see table S1). These findings suggest a flow of information from the lowest to the highest hierarchy with each step enabling greater local processing, possibly supporting increased data integration.

We further tried to distinguish the differences between localized and distributed hierarchies. Distributed hierarchies have high standard deviation of the shell distribution and localized hierarchies have small standard deviation of the shell distribution (see fig. 5). Notably, while most of the edges of the localized hierarchies were mainly self-connections or connections to their distributed partner in the same hierarchy (e.g. distributed to localized middle), the distributed hierarchies displayed more connections to other hierarchies (~15% in distributed subgroups compared to only ~8% in localized subgroups) supporting their role in cross-hierarchy data integration. Moreover, many of these connections were also across similar categories (e.g. distribute middle with distribute high, app. 25%). Furthermore, the distributed and localized subgroups within the same hierarchy displayed a large amount of connections between themselves (~33% of their connections), supporting the fact that they originate from the same hierarchy. The significance of the observed connectivity within and between hierarchies was evaluated using a random permutation test. The results showed that connectivity within each hierarchy is significantly higher (FDR $q<0.0005$) and that connectivity between all hierarchies was significantly lower (FDR $q<0.0005$) than expected according the size of the hierarchies (see Figure 6).

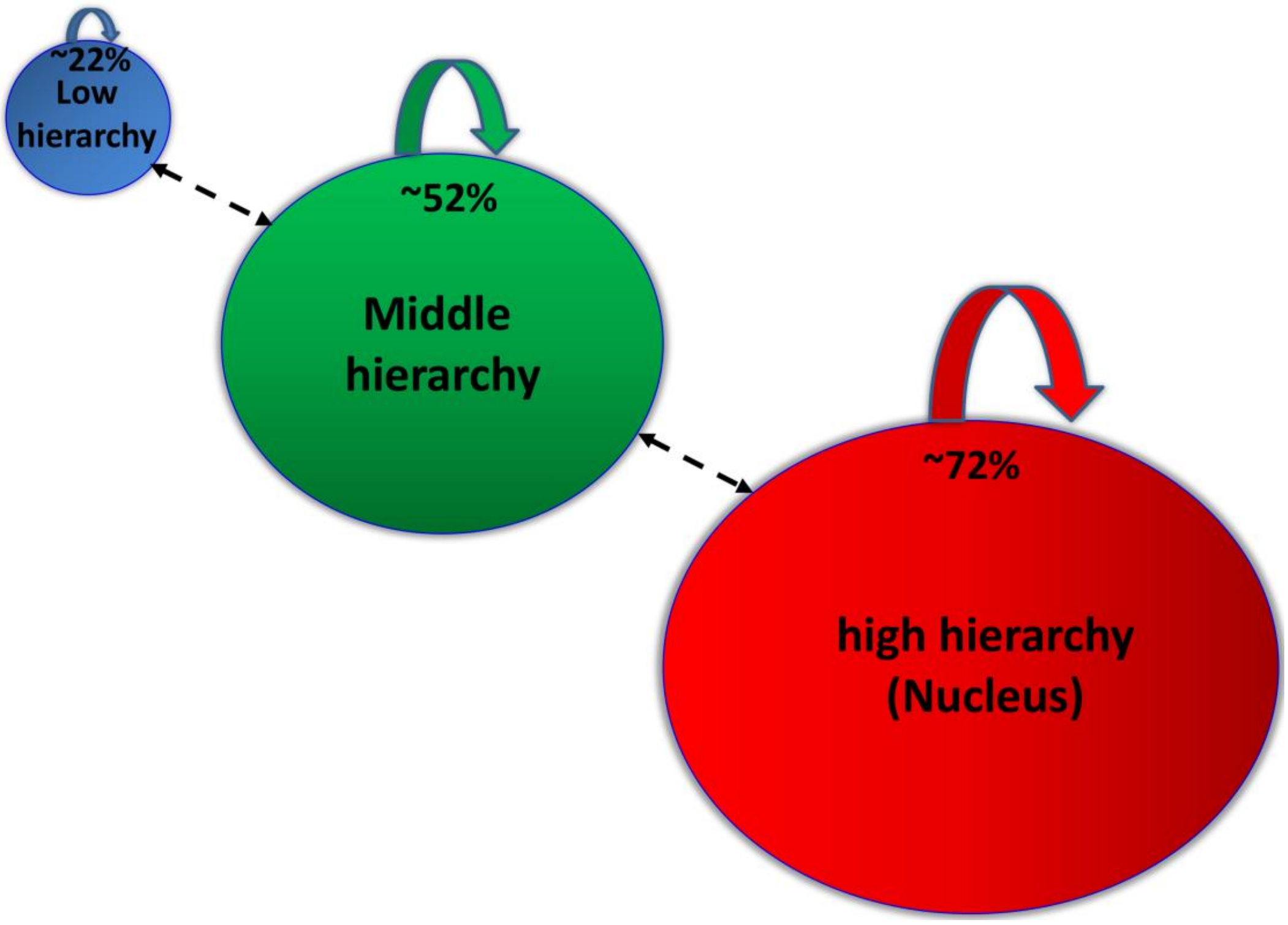

**Fig. 6: Connections between hierarchies.** The size of the hierarchies represents total amount of intra hierarchy connections. Connections within any hierarchy is found to be significantly higher (arrows) and connections between hierarchies was significantly smaller (dash arrows) than expected when taking into account the size of the hierarchies, supporting the modularity nature of every hierarchy. Note the increased self-connections as the hierarchies increase (percent connections are normalized by the total amount of connections in each hierarchy).

## Discussion

In the current study we applied the k-shell decomposition analysis to reveal the global functional organization of the human cortical network. Using this analysis we managed to build a model of cortex topology and connect the structural with the functional level. Our findings indicate that the human cortex is highly connected and efficient, compared to other networks, comprised of a nucleus and a giant component with virtually no isolated nodes. The giant component consists of different degree shells which represent different neighborhoods of connectivity, revealing the global properties of the cortical network. Together with the nucleus, these connectivity shells were categorized into three hierarchies representing an increasing number of regional connections, possibly supporting an increase in data processing and integration within each hierarchy. In accordance, the highest hierarchy was predominantly comprised of left and midline cortical regions (including regions of the default mode network) known to be associated with high-order functions (Northoff *et al.*, 2006). Lastly, this collective of interconnected regions, integrating information throughout the cortex, might allow global properties such as consciousness to emerge.

### *Network properties*

Our findings demonstrate, in accordance with previous work (Achard *et al.*, 2006; Cohen and Havlin, 2010; Ekman *et al.*, 2012) that the cortical network is resilient to small perturbations, highly organized, interconnected and much more efficient compared with a random cortical network or the internet network. K-shell decomposition analysis further proved to be more accurate and provide better resolution of network properties compared to standard methods (e.g. counting degrees, for full details see supplementary material 1).

The two main components of the cortical network, the nucleus and the giant component, both have small world properties though they might serve different roles. A higher clustering coefficient of the giant component alongside short average distance of the nucleus suggest that the majority of local processing takes place within the giant component while the nucleus mainly adds shortcuts and global structures to the network. Indeed, although the nucleus is highly connected, it includes only 50% of all hubs unlike the random nucleus which includes all network hubs (see Fig. S2, S3). These 'peripheral' hubs were located in the giant component and, as previously suggested (Achard *et al.*, 2006), might enable efficient data integration and local information processing. Hubs outside the nucleus might therefore, serve as local processors integrating information from lower shells and transfer it forward to a higher hierarchy, eventually reaching the nucleus (for more information see supplementary material 4).

***Network hierarchies and data integration***

K-shell decomposition analysis reveals that the creation of the giant component entails several critical points. From these critical points we could characterize three major neighborhoods of connectivity or three hierarchies (for more details see supplementary material 3). The regions in the lowest hierarchy appeared to be mostly involved in localized sensory perception (e.g. the fusiform face area and visual "what" stream (Goodale and Milner, 1992)). The different nodes within this hierarchy broadly distributed along the shells which might enable efficient data transfer and processing before sending it to higher hierarchies.

The middle hierarchy is found to be composed of high shells with high degree nodes, though half of them never reached the nucleus, a property that separates these regions from the high hierarchy. Functional regions found in this hierarchy appeared to be involved in high cognitive functions and data integration. For instance, most of the auditory network and regions involved in the integration of audio and visual perception were found in the middle hierarchy. In addition 40% of the executive control network (including right dorsolateral PFC, a crucial region in executive control and working memory (Raz and Buhle, 2006)) and the dorsal visual stream (where\what stream (Goodale and Milner, 1992)) are found in this hierarchy. Broca's area was also located in the middle hierarchy as well as other homologue regions related to language such as Broca and Wernicke homologues.

The high hierarchy contained regions predominantly found in the nucleus. All regions that reached the nucleus across all cortical networks are found in this hierarchy. Unlike other hierarchies, this unique hierarchy is a single, highly interconnected component, which enables high levels of data integration and processing, probably involved in the highest cognitive functions. In accordance, the high hierarchy exhibited the highest amount of self-connections across hierarchies suggesting that it processes data mostly within itself (see Fig. 6). The nucleus (represented by the high hierarchy) has a very strong overlap with the default mode network (81%), in accordance with the result of Hagmann et al (Hagmann *et al.*, 2008), and also with the visual cortex (75%), sensorimotor network (75%) and salience network (71%). The visual dorsal stream and the executive control network also display 60% overlap with the nucleus. Interestingly, all the regions that never appear in the nucleus (across all 6 networks) belong to the right hemisphere, while a strong tendency towards the left hemisphere appeared when examining the nucleus. As mentioned above, all the regions that reached the nucleus are mostly midline or left hemisphere regions. Roughly speaking, the left hemisphere is comprised of high hierarchy regions and the right hemisphere is comprised of middle hierarchy regions (see Fig. 4 and supplementary material 3 and 2).

Looking across hierarchies it's evident that the lowest hierarchy has the smallest amount of connections to other hierarchies and within itself; the middle hierarchy has more connections, almost equally distributed between itself and others; and the high hierarchy has the largest amount of connections, most of them within itself (see Fig. 6). Interestingly, self-connections within each hierarchy were significantly higher and between hierarchies significantly smaller, than expected in random control according the size of the hierarchies. This finding suggests that every hierarchy can be seen as a different module mostly involved in self-processing and only then transfers information to other hierarchies (Bullmore and Sporns, 2009; Hagmann *et al.*, 2008; van den Heuvel and Sporns, 2011). Regarding cross hierarchy connections, it is important to note that most of the connections between middle and high hierarchies occur in their distributed subgroups. This finding suggests that in every hierarchy distributed regions are more involved in data transfer and integration across hierarchies, while localized regions deal more with data processing.

Assuming that data integration requires cross hierarchy connections (the amount of data that a hierarchy receives from other hierarchies – the centrality of the hierarchy (Rubinov and Sporns, 2010)) and data processing depend on interconnected regions (the amount of calculations taking place inside the hierarchy – specialized processing within densely interconnected module (Rubinov and Sporns, 2010)), then data integration and processing seem to increase as we step up in the hierarchies. These findings could therefore suggest a flow of information from the lowest to the highest hierarchy with every hierarchy integrating more data and executing further processing, in line with previous studies and theoretical work (Christoff and Gabrieli, 2000; Gray *et al.*, 2002; Northoff *et al.*, 2006; Damasio, 2000). The low hierarchy receives information, performs specific calculations with its small amount of intra connections and passes the information to the higher hierarchies. The middle hierarchy is further able to integrate more data and locally process more information. At the top, the nucleus receives the most information from all other hierarchies and executes further processing using its dense interconnections, suggesting its vital involvement in data integration within the cortical network.

***The Nucleus as a platform for consciousness***

The regions in the nucleus form one component and constitute the most connected neighborhoods in the cortical network with the highest degrees. In contrast to the giant component, which mostly exhibits local structures (i.e. high clustering coefficient), all the regions in the nucleus form global structures (see supplementary material 4) and densely connect within themselves creating a unique **interconnected collective module** all over the brain**.** The regions and profile of this collective module are consistent with previous work (Collin *et al.*, 2014; Hagmann *et al.*, 2008; van den Heuvel and Sporns, 2011), mostly comprised of posterior medial and parietal regions. Furthermore, in **Hagmann** et al's structural cortical core, 70% of the core's edges were self-connections, similar to our findings within the high hierarchy (72%). In addition, this structural core forms one module and connected with **connector hubs** to all other modules in the network, reflecting our results that the nucleus is a single interconnected module with increased global structures. These findings further suggest that the distributed high hierarchy is composed of such connector hubs, in charge of connecting other hierarchies with the nucleus.

A strong inter-connected nucleus has also been demonstrated by **Sporns et al** suggesting a rich club organization of the human connectome (Collin *et al.*, 2014; van den Heuvel and Sporns, 2011; van den Heuvel *et al.*, 2013). Their results revealed a group of "12 strongly interconnected bihemispheric hub regions, comprising, in the cortex, the precuneus, superior frontal and superior parietal cortex". These six cortical regions were part of our more detailed interconnected nucleus which further includes more regions of the high hierarchy (see Table 1). This interconnected collective module creates one global structure, involving regions from all over the cortex, which may create one global function. Given recent theories that explain consciousness as a complex process of global data integration (Balduzzi and Tononi, 2008; Damasio, 2000; Dehaene and Naccache, 2001; Tononi and Edelman, 1998; Godwin *et al.*, 2015), in particular **G**lobal **W**ork space **T**heory and integrated information theory (Balduzzi and Tononi, 2008; Dehaene and Naccache, 2001; Tononi and Edelman, 1998), one can postulate that such global function could be related to conscious abilities. We therefore suggest that **the global interconnected collective module of the nucleus can** **serve as a platform for consciousness to emerge**. Indeed, all of the regions in the nucleus have been previously correlated to consciousness activities (Achard *et al.*, 2006; Godwin *et al.*, 2015; Goodale and Milner, 1992; Gray *et al.*, 2002; Northoff and Bermpohl, 2004; Northoff *et al.*, 2006; Christoff *et al.*, 2009), especially midline and fronto-parietal regions. The nucleus, receiving the most information from all other hierarchies and integrating it to a unified global function, is therefore a perfect candidate to be the high integrative, global work space region in which consciousness can emerge (for more information see supplementary material 5).

***Study limitations***

Some limitation issues have to be taken into account when interpreting the current results. First, our network is limited only to the cortex; future studies should examine the entire brain network and its influence on the profile of the hierarchies or nucleus. It is possible, for instance, that regions within the low hierarchy (e.g. the fusiform gyrus) might belong to higher hierarchies and are affected by lack of subcortical regions (such as the hippocampus). Lastly, the structural connections of our network were mapped with DSI followed by computational tractography (Hagmann *et al.*, 2008; Hagmann *et al.*, 2007; Hagmann *et al.*, 2003; Schmahmann *et al.*, 2007). Although DSI has been shown to be especially sensitive with regard to detecting fiber crossings (Hagmann *et al.*, 2008; Hagmann *et al.*, 2007; Hagmann *et al.*, 2003; Schmahmann *et al.*, 2007), it must be noted that this method may be influenced by errors in fiber reconstruction, and systematic detection biases.

## Conclusions

The current study used k-shell decomposition analysis in order to reveal the global functional organization of the human cortical network. Consequently, we built a model of human cortex topology and revealed the hierarchical structure of the cortical network. In addition, this analysis proved to be more accurate than standard methods in the characterization of cortical regions and hierarchies. Our findings indicate that the human cortex is highly connected and efficient, compared to other networks, comprised of a nucleus and a giant component with virtually no isolated nodes. The giant component consists of different connectivity shells, which we categorized into three hierarchies representing an increasing number of regional connections. Such a topological model could support an efficient flow of information from the lowest hierarchy to the highest one, with each step enabling more data integration and data processing. At the top, the highest hierarchy (the *global interconnected collective module*) receives information from all previous hierarchies, integrates it into one global function and thus might serve as a platform for consciousness to emerge.

***Acknowledgments***

We would like to thank Mr. Kobi Flax for his crucial role in data analysis, without him this work would not be accomplished! We would also like to thank Dr. Itay Hurvitz for his invaluable comments. We thank the European MULTIPLEX (EU-FET project 317532) project, the Israel Science Foundation, ONR and DTRA for financial support.

*Tables:*

*Table 1: Cortical anatomical regions according to hierarchies*

| Anatomical Region | Side | Function |
|---|---|---|
| **Localized High** | | |
| Paracentral lobule | Mid | SMA - sensorimotor network (Always) |
| Caudal anterior cingulate cortex | L | Salience network (Always) |
| Caudal anterior cingulate cortex | R | Salience\Executive control network (Always) |
| Inferior parietal cortex | L | DMN, Sensorimotor network, Visual dorsal stream (Always) |
| Posterior cingulate cortex | Mid | DMN (Always) |
| Rostral anterior cingulate cortex | Mid | Salience\Executive control network, DMN (Always) |
| Precuneus | Mid | DMN (Always) |
| Isthmus of the cingulate cortex | R | DMN (Always) |
| Pericalcarine cortex | R | Primary visual area |
| Postcentral gyrus | L | Primary somatosensory cortex - Sensorimotor network |
| Superior parietal cortex | L | Executive control, Sensory integration, Sensorimotor network, Visual dorsal stream |
| Supramarginal gyrus | L | Wernicke area, TPJ |
| Bank of the Superior temporal sulcus | L | Visual dorsal stream |
| Cuneus | R | Visual |
| **Distributed high** | | |
| Superior frontal cortex | L | DMN\ Executive\ Salience, Sensorimotor network (Always) |
| Precentral gyrus | L | Primary motor cortex - sensorimotor network (Always) |
| Superior temporal cortex | L | Wernicke ,TPJ, Visual dorsal stream |
| Pericalcarine cortex | L | Primary visual |
| Pars orbitalis | L | Executive control network |
| Middle temporal cortex | L | V5 (Visual dorsal stream), DMN |
| Lateral occipital cortex | L | Primary visual, Visual ventral stream |
| Isthmus of the cingulate cortex | L | DMN |
| Cuneus | L | Visual |
| Rostral middle frontal cortex | L | Executive control network, DMN |
| Superior parietal cortex | R | Executive, sensory integration, Sensorimotor network, Visual dorsal stream |
| Superior frontal cortex | R | DMN\ Executive\ Salience\ Sensorimotor network |
| Postcentral gyrus | R | Primary somatosensory cortex - Sensorimotor network |
| Lingual gyrus | R | Visual |
| **Localized middle** | | |
| Inferior parietal cortex | R | DMN, Sensorimotor network, Visual dorsal stream (Never) |
| Caudal middle frontal cortex | R | Executive control network, Sensorimotor network (Never) |
| Bank of the superior temporal sulcus | R | Visual dorsal stream (Never) |
| Supramarginal gyrus | R | Wernicke homologue, TPJ (Never) |
| Superior temporal cortex | R | Wernicke homologue, TPJ, Visual dorsal stream (Never) |
| Frontal pole | R | Executive control network |
| Frontal pole | L | Salience and executive control networks |
| Medial orbitofrontal cortex | R | Stimulus-reward associations |
| **Distributed middle** | | |
| Pars triangularis | R | Broca homologue (Never) |
| Pars triangularis | L | Broca |
| Middle temporal cortex | R | V5 (Visual dorsal stream), DMN (Never) |
| Pars opercularis | R | Broca homologue (Never) |
| Pars opercularis | L | Broca |
| Inferior temporal cortex | R | Visual association, Visual ventral stream (Never) |
| Inferior temporal cortex | L | Visual association, Visual ventral stream |
| Rostral middle frontal cortex | R | Salience and executive control networks (Never) |
| Pars orbitalis | R | Salience and executive control networks (Never) |
| Transverse temporal cortex | R | Primary auditory cortex (Never) |
| Temporal pole | L | Salience network |
| Lateral orbitofrontal cortex | L+R | Stimulus-reward associations |
| Medial orbitofrontal cortex | L | Stimulus-reward associations |
| Precentral gyrus | R | Primary motor cortex - Sensorimotor network |
| Caudal middle frontal cortex | L | Executive control network, DMN, Sensorimotor network |
| Lateral occipital cortex | R | Primary visual, Visual ventral stream |
| **Low** | | |
| Temporal pole | R | Salience network (Never) |
| Parahippocampal cortex | R | Hippocampal support, Visual ventral stream (Never) |
| Parahippocampal cortex | L | Hippocampal support, Visual ventral stream |
| Fusiform gyrus | R | Face recognition, Visual ventral stream (Never) |
| Fusiform gyrus | L | Face recognition, Visual ventral stream |
| Entorhinal cortex | R | Hippocampal support, Visual ventral stream (Never) |
| Entorhinal cortex | L | Hippocampal support, Visual ventral stream |
| Lingual gyrus | L | Visual association |

DMN= Default mode network, TPJ=temporal parietal junction. Always= region that always reaches the nucleus for all networks, Never= region that never reaches the nucleus for all networks.

*Table 2: Laterality effects*

| Anatomical region | Left | Right |
|---|---|---|
| Precentral gyrus (primary motor cortex) | High (always) | Middle |
| Inferior parietal | High (always) | Middle (never) |
| Supra marginal gyrus (Wernicke area,TPJ) | High | Middle (never) |
| Superior temporal (Wernicke area ,TPJ) | High | Middle (never) |
| Lateral occipital cortex (primary visual) | High | Middle |
| Lingual gyrus (visual association) | Low | High |
| Bank of the superior temporal sulcus (vision) | High | Middle (never) |
| Pars Orbitalis (executive control network) | High | Middle (never) |
| Middle temporal (V5, DMN) | High | Middle (never) |
| Rostral middle frontal cortex (executive control network, DMN) | High | Middle (never) |
| Superior frontal cortex | High (always) | High |
| Caudal middle frontal cortex (executive control network, DMN) | Middle | Middle (never) |
| Inferior temporal cortex (visual association) | Middle | Middle (never) |
| Pars triangularis (Broca homologue) | Middle | Middle (never) |
| Pars opercularis (Broca homologue) | Middle | Middle (never) |
| Temporal pole (salience network) | Middle | Low (never) |
| Parahippocampal cortex | Low | Low (never) |
| Fusiform gyrus | Low | Low (never) |
| Entorhinal cortex | Low | Low (never) |
| **Functional Networks** | | |
| Dorsal stream (where stream) | 100% high | 80% middle (80% never) |
| Ventral stream (what stream) | 60% low | 60% low (80% never) |
| Auditory network | 100% high | 100% middle (100% never) |
| Executive control network | 77% high | 55% middle |
| Default mode network | 89% high (55% always) | 71% high (57% always) |
| Salience network | 60% (always) | 40% (always) |
| Sensorimotor network | 83% (always) | 17% (always) |

DMN= Default mode network, TPJ=temporal parietal junction. Always= region that always reaches the nucleus for all networks, Never= region that never reaches the nucleus for all networks.